\newif\ifonecol

\onecoltrue    

\ifonecol
\documentclass[11pt,draftcls,onecolumn,peerreview]{IEEEtran}
\else
\documentclass[journal]{IEEEtran}
\fi

\usepackage{amssymb,amsmath,xspace,intmacros,subfigure,amsfonts}
\usepackage[final]{graphicx}

\newcommand{\xb}{{\textbf{x}}}

\newcommand{\Db}{{\textbf{D}}}
\newcommand{\Gb}{{\textbf{G}}}
\newcommand{\db}{{\textbf{d}}}

\newcommand{\Jb}{{\textbf{J}}}
\newcommand{\eb}{{\textbf{e}}}

\newcommand{\Ib}{{\textbf{I}}}

\newcommand{\wb}{{\textbf{w}}}

\newcommand{\yb}{{\textbf{y}}}

\newcommand\ie{i.e.,\xspace}

\newcommand{\lzero}{$\ell^{0}$}
\newcommand{\lone}{$\ell^{1}$}

\newcounter{examplecounter}
\renewcommand{\theexamplecounter}{\arabic{examplecounter}}

\newcommand{\inst}[1]{\unskip${^{#1}}$}

\begin{document}


\title{Bayesian Cram\'{e}r-Rao Bound for Noisy Non-Blind and Blind Compressed Sensing }
\author{H.~Zayyani \inst{1}, M.~Babaie-Zadeh\inst{1}~\IEEEmembership{Member},
 and C.~Jutten \inst{2}~\IEEEmembership{Fellow}\\
 \ifonecol
 EDICS:SAS-ICAB or SAS-STAT
 \fi
 \thanks{$^1$Electrical engineering department, Sharif university of
 technology, Tehran, Iran.}
 \thanks{$^2$GIPSA-LAB, Grenoble, and Institut Universitaire de France, France.}
 \thanks{This work has been partially funded by Iran NSF (INSF)
under contract number 86/994, by Iran Telecom Research Center
(ITRC), and also by center for International Research and
Collaboration (ISMO) and French embassy in Tehran in the framework
of a GundiShapour collaboration program.}
 \ifonecol
 \thanks{First author: Hadi Zayyani, email: {\tt zayyani@ee.sharif.edu}, Tel: +98 21 66164125, Fax: +98 21 66023261.}
 \thanks{Second author: Masoud Babaie-zadeh, email: {\tt mbzadeh@sharif.edu}, Tel: +98 21 66165925, Fax: +98 21 66023261.}
 \thanks{Fourth author: Christian Jutten, email: {\tt  christian.jutten@gipsa-lab.grenoble-inp.fr}, Tel: +33 (0)4 76574351, Fax: +33 (0)4 76574790.}
 \fi
 }
\markboth{IEEE Signal Processing Letters, Vol. XX, No. Y, Month
2009}{Zayyani, Babaie-Zadeh and Jutten: Sparse Component Analysis
in Presence of Noise by a New EM Algorithm}

\maketitle

\begin{abstract}
In this paper, we address the theoretical limitations in
reconstructing sparse signals (in a known complete basis) using
compressed sensing framework. We also divide the CS to non-blind
and blind cases. Then, we compute the Bayesian Cramer-Rao bound
for estimating the sparse coefficients while the measurement
matrix elements are independent zero mean random variables.
Simulation results show a large gap between the lower bound and
the performance of the practical algorithms when the number of
measurements are low.

\end{abstract}

\section{Introduction}
\label{sec:intro} Compressed Sensing or Compressive Sampling
(CS)~\cite{CandT06},~\cite{Dono06} is an emerging field in signal
processing. The theory of CS suggests to use only a few random
linear measurement of a sparse signal (in a basis) for
reconstructing the original signal. The mathematical model of
noise free CS is:
\begin{equation}
\yb=\boldsymbol{\Phi}\xb
\end{equation}
where $\xb=\boldsymbol{\Psi}\wb$ is the original signal with
length $m$ and is sparse in the basis $\boldsymbol{\Psi}$ ($\ie
||\wb||_0<K$ and $K$ is defined as sparsity level) and
$\boldsymbol{\Phi}$ is an $n\times m$ random measurement matrix
where $n<m$. For near perfect recovery, in addition to the signal
sparsity, the incoherence of the random measurement matrix
$\boldsymbol{\Phi}$ with the basis $\boldsymbol{\Psi}$ is needed.
The incoherence is satisfied with high probability for some types
of random matrices such as i.i.d Gaussian elements or i.i.d
Bernoulli $\pm1$ elements. Recent theoretical results show that
under these two conditions (sparsity and incoherence), the
original signal can be recovered from only a few linear
measurements of the signal within a controllable error, even in
the case of noisy measurements
\cite{CandT06},~\cite{Dono06},~\cite{HaupN06},~\cite{AkcaT07}.

In \cite{HaupN06}, some error bounds are introduced for
reconstructing the original sparse (or compressible) signal in the
noisy CS framework. In \cite{AkcaT07}, the performance limits of
noisy CS is investigated by definition of some performance metrics
which are of Shannon Theoretic spirit. \cite{CandT06} considers
the no noise CS and finds an upper bound on reconstruction error
in terms of Mean Square Error (MSE) only for \lone-minimization
recovery algorithm. But, \cite{HaupN06} finds some upper bounds in
the noisy CS and for general recovery algorithms. \cite{AkcaT07}
is also investigated its own decoder which is derived based on
joint typicality. Moreover, some information theoretic bounds are
derived in \cite{AeroZS07}.


In this paper, we derive a Bayesian Cramer-Rao Bound (BCRB)
(\cite{Vant68},~\cite{TichMN98}), which is a lower bound, for
noisy CS by a statistical view to the CS problem. This BCRB bounds
the performance of any parametric estimator (whether biased or
unbiased) of the sparse coefficient vector in terms of mean square
estimation error \cite{Vant68},~\cite{TichMN98}. We also introduce
the notion of blind CS in contrast to the traditional CS to whom
we refer on the non-blind CS. We compute BCRB for both non-blind
and blind CS, where in the latter, we do not know the measurement
matrix in advance. In a related direction of research, a CRB is
obtained for mixing matrix estimation in Sparse Component Analysis
(SCA) \cite{ZayyBHJ08}.

\section{Non-blind and blind noisy CS}
Consider the noisy CS problem:
\begin{equation}
\label{eq: ncs}
\yb=\boldsymbol{\Phi}\boldsymbol{\Psi}\wb+\eb=\Db\wb+\eb
\end{equation}
where $\Db=\boldsymbol{\Phi}\boldsymbol{\Psi}$, $\wb$ is a sparse
vector and $\eb$ is a Gaussian zero-mean noise vector with the
covariance $\sigma^2_e\Ib$. In CS framework, we want to estimate
$\wb$, from which, $\xb=\boldsymbol{\Psi}\wb$ can be reconstructed
from the measurement vector $\yb$.

We nominate the traditional CS problem as non-blind CS since we
know the basis $\boldsymbol{\Psi}$ and the measurement matrix
$\boldsymbol{\Phi}$ and hence $\Db$ in advance. In some cases, we
have no prior information about the signals in addition to their
sparsity. As such, we do not know the basis $\boldsymbol{\Psi}$,
in which the signals are sparse. One application is a blind
interceptor who intercepts the signals. The only information is
that the signals have been received are sparse in some unknown
domain. In these cases, we nominate the problem as blind CS which
is inspired from the well known problem of Blind Source Separation
(BSS). As such, each measurement will be:
\begin{equation}
\label{eq: blind1}
y=\boldsymbol{\phi}^T\boldsymbol{\Psi}\wb+\eb=\db^T\wb+\eb
\end{equation}
where $\boldsymbol{\phi}^T$ is the random measurement vector and a
row of $\boldsymbol{\Phi}$) and
$\db^T=\boldsymbol{\phi}^T\boldsymbol{\Psi}$ is the corresponding
row in $\Db$ and an unknown random vector.


\section{Bayesian Cramer-Rao Bound}
\label{sec: CRLB}

The Posterior Cramer-Rao Bound (PCRB) or Bayesian Cramer-Rao Bound
(BCRB) of a vector of parameters $\boldsymbol{\theta}$ estimated
from data vector $\yb$ is the inverse of the Fisher information
matrix, and bounds the estimation error in the following form
\cite{TichMN98}:
\begin{equation}
\label{eq: BCRB}
E\left[(\boldsymbol{\theta}-\hat{\boldsymbol{\theta}})(\boldsymbol{\theta}-\hat{\boldsymbol{\theta}})^T\right]\ge\Jb^{-1}
\end{equation}
where $\boldsymbol{\hat{\theta}}$ is the estimate of
$\boldsymbol{\theta}$ and $\Jb$ is the Fisher information matrix
with the elements \cite{TichMN98}:
\begin{equation}
J_{ij}=E_{\yb,\boldsymbol{\theta}}\left[-\frac{\partial^2\log
p(\yb,\boldsymbol{\theta})}{\partial\theta_i\partial\theta_j}
\right],
\end{equation}
where $p(\yb,\boldsymbol{\theta})$ is the joint probability
between the observations and the parameters. Unlike CRB, the BCRB
(\ref{eq: BCRB}) is satisfied for any estimator (even for biased
estimators) under some mild conditions \cite{Vant68},
\cite{TichMN98} which we assume that are fulfilled in our problem.
Using Bayes rule, the Fisher information matrix can be decomposed
into two matrices \cite{TichMN98}:
\begin{equation}
\Jb=\Jb_{D}+\Jb_P,
\end{equation}
where $\Jb_{D}$ represents data information matrix and $\Jb_P$
represents prior information matrix which their elements are
\cite{TichMN98}:
\begin{equation}
J_{D_{ij}}\triangleq
E_{\yb,\boldsymbol{\theta}}\left[-\frac{\partial^2\log
p(\yb|\boldsymbol{\theta})}{\partial\theta_i\partial\theta_j}\right]=E_{\boldsymbol{\theta}}(J_{s_{ij}})
\end{equation}
\begin{equation}
\label{eq: pfim} J _{P_{ij}}\triangleq
E_{\boldsymbol{\theta}}\left[-\frac{\partial^2\log
p(\boldsymbol{\theta})}{\partial\theta_i\partial\theta_j}\right]
\end{equation}
where $\Jb_{s}\triangleq
E_{\yb|\boldsymbol{\theta}}[-\frac{\partial^2\log
p(\yb|\boldsymbol{\theta})}{\partial\theta_i\partial\theta_j}]$ is
the standard Fisher information matrix \cite{Kay93} and
$p(\boldsymbol{\theta})$ is the prior distribution of the
parameter vector.

In this paper, we use this BCRB for our problem because we have a
sparse prior information about the parameter which is estimated.
We compute BCRB for two blind and non-blind cases.

\subsection{Computing BCRB in non-blind CS}
In the non-blind CS case, the matrices $\boldsymbol{\Phi}$ and
$\boldsymbol{\Psi}$ are assumed to be known and
$\boldsymbol{\Phi}$ is a random matrix while $\boldsymbol{\Psi}$
is a fixed basis matrix. Similar to \cite{WiesEY08}, since
$\boldsymbol{\Phi}$ is assumed to be known and random,
$\boldsymbol{\Phi}$ can be added as an additional observation.
Hence, the data information matrix elements $\Jb_{D_{ij}}$ from
model (\ref{eq: ncs}) are of the form:
\begin{equation}
\label{eq: expec}
J_{D_{ij}}=E_{\yb,\wb,\boldsymbol{\Phi}}\left[-\frac{\partial^2\log
p(\yb,\boldsymbol{\Phi}|\wb)}{\partial\w_i\partial\w_j}\right].
\end{equation}
since
$p(\yb,\boldsymbol{\Phi}|\wb)=p(\boldsymbol{\Phi})p(\yb|\boldsymbol{\Phi},\wb)$,
$p(\boldsymbol{\Phi})$ is independent of $\wb$ and
$p(\yb|\boldsymbol{\Phi},\wb)=(2\pi\sigma^2_e)^{\frac{-n}{2}}\exp(\frac{-1}{2\sigma^2_e}||\yb-\Db\wb||^2_2)$,
we can write $\frac{\partial \log
p(\yb,\boldsymbol{\Phi}|\wb)}{\partial\wb}=\frac{-1}{2\sigma^2_e}(-2\yb^T\Db+2\Db^T\Db\wb)$.
So, we have $\frac{\partial \log
p(\yb,\boldsymbol{\Phi}|\wb)}{\partial
w_i}=\frac{1}{\sigma^2_e}(\yb^T\Db)_i-\frac{1}{\sigma^2_e}\sum_{r=1}^m
g_{ir}w_r$ where $g_{ij}$ denotes the elements of the matrix
$\Gb=\Db^T\Db$. Hence, we have $\frac{\partial^2\log
p(\yb,\boldsymbol{\Phi}|\wb)}{\partial w_i\partial
w_j}=\frac{-1}{\sigma^2_e}g_{ij}$. So, the expectation (\ref{eq:
expec}) will be $J_{D_{ij}}=E_{\yb,\wb,\boldsymbol{\Phi}}\left[
\frac{1}{\sigma^2_e}g_{ij}\right]=\frac{1}{\sigma^2_e}E_{\boldsymbol{\Phi}}\{g_{ij}\}=J_{D_{ij}}=\frac{1}{\sigma^2_e}\sum_{r=1}^n
E_{\boldsymbol{\Phi}}\{d_{ri}d_{rj}\} $. Some simple manipulations
show that under assumption that the elements of
$\boldsymbol{\Phi}$ are zero mean and independent random
variables, the data information matrix will be:
\begin{equation}
\Jb_{D}=n\frac{\sigma^2_r}{\sigma^2_e}\boldsymbol{\Psi}^T\boldsymbol{\Psi}
\end{equation}
where $\sigma^2_r=E(\phi^2_{ij})$ is the variance of the random
measurement matrix elements. If $\Psi$ is an orthonormal basis
then $\boldsymbol{\Psi}^T\boldsymbol{\Psi}=\Ib$ and hence
$\Jb_{D}=n\frac{\sigma^2_r}{\sigma^2_e}\Ib$.

To compute the prior information matrix $\Jb_{P}$ from (\ref{eq:
pfim}), we should assume a sparse prior distribution for our
parameter vector elements $w_i$. Similarly to \cite{WipfR04}, we
assume $w_i$'s are independent and have a parameterized Gaussian
distribution:
\begin{equation}
\label{eq: pr}
p(w_i)=\frac{1}{\sigma_i\sqrt{2\pi}}\exp(-\frac{w^2_i}{2\sigma^2_i}),
\end{equation}
In (\ref{eq: pr}), the variance $\sigma^2_i$ enforce the sparsity
of the corresponding coefficient: a small variance means that the
coefficient is inactive and a large value means the activity of
the coefficient. It can be easily seen that in this case, the
prior information matrix is $\Jb_P=\diag(\frac{1}{\sigma^2_i})$.
Finally, for orthonormal bases for $\boldsymbol{\Psi}$ and for
prior distribution (\ref{eq: pr}), the BCRB results in:
\begin{equation}
\label{eq: NBCRB} E\left[(w_i-\hat{w_i})^2\right]\ge
\left(n\frac{\sigma^2_r}{\sigma^2_e}+\frac{1}{\sigma^2_i}\right)^{-1}.
\end{equation}

\subsection{Computing BCRB in blind CS}
In the blind CS case, the matrix $\boldsymbol{\Psi}$ is not known
in advance and hence the elements of matrix $\Db$ are random and
unknown with zero mean. If we restrict ourselves to Gaussian
measurements matrix elements ($\phi_{ij}$ is a zero-mean Gaussian)
then different measurement samples of $y$ are also Gaussian and
independent of each other. Hence, we can compute the data
information matrix from only one measurement (\ref{eq: blind1}).
Then, the information matrix elements
$J_{D_{ij}}=E_{y,\wb}\left[-\frac{\partial^2\log
p(y|\wb)}{\partial\w_i\partial\w_j}\right]$ will be equal to
(refer to \cite{Kay93}):
\begin{equation}
\label{eq: bbcrb} J_{D_{ij}}=E_{y,\wb}\left[\frac{\partial\log
p(y|\wb)}{\partial\w_i}\frac{\partial\log p(y|\wb)}{\partial\w_j}
\right].
\end{equation}
If the elements of $\boldsymbol{\phi}$ are assumed to be random
with a Gaussian distribution of zero mean and variance
$\sigma^2_r$ and the columns of the basis matrix
$\boldsymbol{\Psi}$ have unit norms, then:
\begin{equation}
\label{eq: blind}
p(y|\wb)=\frac{1}{\sqrt{2\pi\sigma^2(\wb)}}\exp(-\frac{y^2}{2\sigma^2(\wb)})
\end{equation}
where $\sigma^2(\wb)\triangleq\sigma^2_e+\sigma^2_r||\wb||^2_2$.
Simple manipulations show:
\begin{equation}
\frac{\partial\log p(y|\wb)}{\partial
w_i}=-\frac{w_i\sigma^2_r}{\sigma^4(\wb)}\left(\sigma^2(\wb)-y^2\right)
\end{equation}
and from (\ref{eq: bbcrb}) we should compute:
\begin{equation}
J_{D_{ij}}=\sigma^4_r\int_{\wb}\frac{w_iw_j}{\sigma^8(\wb)}
\left[\int_{y}(\sigma^2(\wb)-y^2)^2p(y|\wb)dy\right]p(\wb)d\wb
\end{equation}
where the internal integral is
$\int_{y}(\sigma^2(\wb)-y^2)^2p(y|\wb)dy=m_4-2\sigma^2(\wb)m_2+\sigma^4(\wb)$
in which $m_2$ and $m_4$ are the second and fourth order moments
equal to $m_2=\sigma^2(\wb)$ and $m_4=3\sigma^4(\wb)$. So, we have
$\int_{y}(\sigma^2(\wb)-y^2)^2p(y|\wb)dy=2\sigma^4(\wb)$ and then:
\begin{equation}
J_{D_{ij}}=2\sigma^4_r\int_{\wb}\frac{w_iw_j}{\sigma^4(\wb)}p(\wb)d\wb
\end{equation}
where the off diagonal terms are zeros $J_{D_{ij}}=0,j\neq i$
because the integrand is an odd function. The diagonal terms are:
\begin{equation}
J_{D_{ii}}=2\sigma^4_r\int_{\wb}\frac{w^2_i}{(\sigma^2_e+\sigma^2_r||\wb||^2_2)^2}p(\wb)d\wb
\end{equation}

Following Appendix~\ref{app1}, the diagonal elements are
simplified as:
\begin{equation}
\label{eq: final}
J_{D_{ii}}=\frac{2\sigma^2_r}{m}\left(A_1-\sigma^2_eA_2\right)
\end{equation}
where $A_1$ and $A_2$ are defined and calculated in
Appendix~\ref{app1}.

The prior information matrix for BG distribution
$p(w_i)=p\delta(w_i)+(1-p)\frac{1}{\sigma\sqrt{2\pi}}\exp(-\frac{w^2_i}{2\sigma^2})$
is calculated in Appendix~\ref{app2}:
\begin{equation}
\Jb_P=\frac{1-p}{\sigma^2}\Ib
\end{equation}
Finally, the Blind BCRB is calculated as:
\begin{equation}
\label{eq: BBCRB} E\left[(w_i-\hat{w_i})^2\right]\ge
\left(2\frac{\sigma^2_r}{m}\left(A_1-\sigma^2_eA_2\right)+\frac{1-p}{\sigma^2}\right)^{-1}
\end{equation}

\section{Simulation results}
\label{sec: simresult} 
\begin{figure}[tb]
\begin{center}
\includegraphics[width=7cm]{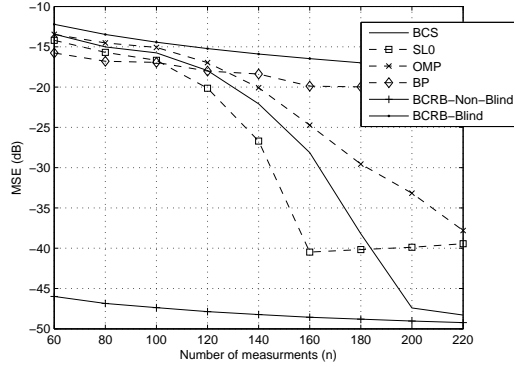}
\end{center}
\caption{ MSE versus number of measurements for a sparse signal in
the time domain ($\boldsymbol{\Psi}=\Ib$) with length $m=512$ and
with the BG distribution with parameters $p=0.9$, $\sigma_1=0.5$
and $\sigma_2=0$. Measurement matrix elements are unit variance
Gaussian random variables.}
\label{fig1}
\end{figure}

In this section, we compare the CRB's with
the results of some of the state-of-the-art algorithms for signal
reconstruction in CS. In our simulations, we used sparse signals
with the length $m=512$ in the time domain where
$\boldsymbol{\Psi}=\Ib$. We used a BG distribution with the
probability of being nonzero equal to $1-p=0.1$ and the variance
for nonzero coefficients is equal to $\sigma^2=(0.5)^2$. So, in
average there were 51 active coefficients. We used a Gaussian
random measurement matrix with elements drawn from zero mean
Gaussian distribution with variance equal to $\sigma^2_r=1$. The
number of measurements are varied between 60 to 200. We computed
the Mean Square Error (MSE) for sparse coefficient vector over 100
different runs of the experiment:
\begin{equation}
\mbox{MSE}\triangleq10\log_{10}\left(\frac{1}{100}\sum_{r=1}^{100}||\wb_r-\hat{\wb}_r||^2_2\right)
\end{equation}
where $r$ is the experiment index. We compared this measure for
various algorithms with the average value of BCRB for non-blind
case which is equal to $\frac{1}{m}\mbox{trace}(\Jb^{-1})$. The
algorithms used for our simulation are Orthogonal Matching Pursuit
(OMP) \cite{PatiRK93}, Basis Pursuit (BP) \cite{ChenDS98},
Bayesian Compressive Sampling (BCS) \cite{JiXC08} and Smoothed-L0
(SL0) \footnote{We used the OMP code from
http://sparselab.stanford.edu with 50 iterations, the BP code from
http://www.acm.caltech.edu/l1magic/l1eq-pd.m with pdtol=1e-6 and
its default parameters, the BCS code from
http://people.ee.duke.edu/\~{}lihan/cs with its default parameters
and the  SL0 code from http://ee.sharif.edu/\~{}SLzero with
parameters \mbox{sigma-min}=0.001 and
\mbox{sigma-decrease-factor}=0.9.} \cite{MohiBJ08}. We also
computed the BCRB for blind case (\ref{eq: BBCRB}) to compare the
BCRB's in both blind and non-blind case. Figure~\ref{fig1} shows
the results of the simulation. It can be seen that in the low
number of measurements, there is a gap between the BCRB and the
performance of algorithms while one of the algorithms
approximately reaches the BCRB for large number of measurements.
Moreover, the difference between the BCRB's for the non-blind and
blind cases are very large. It shows that the blind case needs
much more linear measurements than the non-blind case.

To verify the approximation $D_1\approx 0$ and $D_2\approx 0$
(refer to Appendix~\ref{app2}), we calculated the integrals
numerically with parameters $p=0.9$ and $\sigma=1$. When
$\sigma_0=10^{-5}$ then $D_1=4.7990\times 10^{-25}$ and
$D_2=2.7673\times 10^{-19}$. It shows that our approximations are
true for sufficiently small value of $\sigma_0$.

\section{Conclusions}
In this paper, the CS problem is divided into non-blind and blind
cases and the Bayesian Cramer-Rao bound for estimating the sparse
vector of the signal was calculated in the two cases. The
simulation results show a large gap between the lower bound and
the performance of the practical algorithms when the number of
measurements are low. There was also a large gap between the BCRB
in both non-blind and blind cases. It also shows that in the blind
CS framework, much more blind linear measurements of the sparse
signal are needed for perfect recovery of the signal.


\appendices

\section{Computing the integral}
\label{app1} Let define
$I_i=\int_{\wb}\frac{w^2_i}{(\sigma^2_e+\sigma^2_r||\wb||^2_2)^2}p(\wb)d\wb$
and assume an equal prior distribution for all coefficients $w_i$,
then all $I_i$'s are the same because of the symmetry of the
integral. So, we can add all the integrals and write:
\begin{equation}
\begin{split}
m\sigma^2_rI_i=\int_{\wb}\frac{\sigma^2_r||\wb||^2_2}{(\sigma^2_e+\sigma^2_r||\wb||^2_2)^2}p(\wb)d\wb=\\
\int_{\wb}\frac{p(\wb)}{(\sigma^2_e+\sigma^2_r||\wb||^2_2)}d\wb-\sigma^2_e\int_{\wb}\frac{p(\wb)}{(\sigma^2_e+\sigma^2_r||\wb||^2_2)^2}d\wb
\end{split}
\end{equation}
Then, if we nominate the two above integrals as
$A_1=\int_{\wb}\frac{p(\wb)}{(\sigma^2_e+\sigma^2_r||\wb||^2_2)}d\wb$
and
$A_2=\int_{\wb}\frac{p(\wb)}{(\sigma^2_e+\sigma^2_r||\wb||^2_2)^2}d\wb$,
the integral $I_i$ is computed as
$I_i=\frac{1}{m\sigma^2_r}\left(A_1-\sigma^2_eA_2\right)$. To
compute $A_1$ and $A_2$, we approximate the joint probability
distribution of coefficients as:
\begin{equation}
\begin{split}
p(\wb)=\prod_{i=1}^mp(w_i)\approx p^m\prod_{i=1}^m\delta(w_i)+\\
p^{m-1}(1-p)\sum_{r=1}^m\frac{\prod_{i=1,i\neq
r}^m\delta(w_i)}{\sigma\sqrt{2\pi}}\exp\Big(-\frac{w^2_i}{2\sigma^2_2}\Big)
\end{split}
\end{equation}
This approximation is based on the assumption that the value of
$(1-p)$ which is the activity probability is very small and so we
can neglect the higher order powers of $(1-p)$. By this
approximation, the two integrals will be approximately:
\begin{displaymath}
\label{eq: A1}
A_1=\frac{p^m}{\sigma^2_e}+\frac{mp^{m-1}(1-p)}{\sigma\sqrt{2\pi}}B_1
\end{displaymath}
\begin{displaymath}
\label{eq: A2}
A_2=\frac{p^m}{\sigma^2_4}+\frac{mp^{m-1}(1-p)}{\sigma\sqrt{2\pi}}B_2
\end{displaymath}
where the two integrals are
$B_1=\int_{w}\frac{\exp(-\frac{w^2}{2\sigma^2})}{(\sigma^2_e+\sigma^2_rw)}dw$
and
$B_2=\int_{w}\frac{\exp(-\frac{w^2}{2\sigma^2})}{(\sigma^2_e+\sigma^2_rw)^2}dw$.
By change of variable $x=\frac{w}{\sigma\sqrt{2}}$, the two
integrals are equal to:
\begin{displaymath}
\label{eq: B1}
B_1=\frac{1}{\sqrt{2}\sigma\sigma^2_r}\int_{-\infty}^{+\infty}\frac{\exp(-x^2)}{a^2+x^2}dx=\frac{1}{\sqrt{2}\sigma\sigma^2_r}C_1
\end{displaymath}
\begin{displaymath}
\label{eq: B2}
B_2=\frac{1}{2\sqrt{2}\sigma^3\sigma^4_r}\int_{-\infty}^{+\infty}\frac{\exp(-x^2)}{(a^2+x^2)^2}dx=\frac{1}{2\sqrt{2}\sigma^3\sigma^4_r}C_2
\end{displaymath}
where $a^2=\frac{\sigma^2_e}{2\sigma^2_r\sigma^2}$. The above
integrals are equal to\footnote{We used Maple software to compute
the integrals analytically.}:
\begin{displaymath}
C_1=\frac{\pi}{a}\exp(a^2)\left[1-\mbox{erf}(a)\right]
\end{displaymath}
\begin{displaymath}
C_2=\frac{\pi\exp(a^2)}{2a^3}\left[1-2a^2+\frac{2\sqrt{\pi}}{\pi\exp(a^2)}-\mbox{erf}(a)+2a^2\mbox{erf}(a)\right]
\end{displaymath}
where $\mbox{erf}(x)$ is the error function, defined as
$\mbox{erf}(x)\triangleq\frac{2}{\sqrt{\pi}}\int_0^x
\exp(-t^2)dt$.

\section{Prior information matrix for BG distribution}
\label{app2} Since the coefficients $w_i$'s are independent, the
off diagonal terms $J_{P_{ij}},i\neq j$ are zero. Because of the
independence of $w_i$'s, we can write
$J_{P_{ii}}=E_{w_i}\{-\frac{\partial^2\log p(w_i)}{\partial^2 w_i}
\}$. To calculate this term, we use a Gaussian distribution with
small variance $\sigma^2_0$ instead of delta function
$\delta(w_i)$. So, the prior is:
\begin{equation}
p(w_i)=A\exp\Big(-\frac{w^2_i}{2\sigma^2_0}\Big)+B\exp\Big(-\frac{w^2_i}{2\sigma^2}\Big)
\end{equation}
where $A=\frac{p}{\sigma_0\sqrt{2\pi}}$,
$B=\frac{1-p}{\sigma\sqrt{2\pi}}$ and $\sigma_0\rightarrow 0$. The
partial derivative can be calculated as:
\begin{equation}
\frac{\partial^2\log p(w_i)}{\partial
w^2_i}=\frac{1}{p(w_i)}\frac{\partial^2p(w_i)}{\partial
w^2_i}-\frac{1}{p^2(w_i)}\Big(\frac{\partial p(w_i)}{\partial
w_i}\Big)^2
\end{equation}
Hence, we have:
\begin{equation}
\label{eq: jbp}
J_{P_{ii}}=-\int_{-\infty}^{+\infty}\frac{\partial^2p(w_i)}{\partial
w^2_i}dw_i+\int_{-\infty}^{+\infty}\frac{1}{p(w_i)}\Big(\frac{\partial
p(w_i)}{\partial w_i}\Big)^2dw_i
\end{equation}
To compute the above integrals, the partial derivatives are
$\frac{\partial p(w_i)}{\partial
w_i}=-\frac{Aw_i}{\sigma^2_0}\exp(-\frac{w^2_i}{2\sigma^2_0})-\frac{Bw_i}{\sigma^2}\exp(-\frac{w^2_i}{2\sigma^2})$
and $\frac{\partial^2 p(w_i)}{\partial
w^2_i}=-\frac{A}{\sigma^2_0}\exp(-\frac{w^2_i}{2\sigma^2_0})+\frac{Aw^2_i}{\sigma^4_0}\exp(-\frac{w^2_i}{2\sigma^2_0})
-\frac{B}{\sigma^2}\exp(-\frac{w^2_i}{2\sigma^2})+\frac{Bw^2_i}{\sigma^4}\exp(-\frac{w^2_i}{2\sigma^2})$.
Simple calculations show that
$\int\frac{\partial^2p(w_i)}{\partial w^2_i}dw_i=0$ and hence:
\begin{equation}
J_{P_{ii}}=\int_{-\infty}^{+\infty}\frac{[-\frac{Aw_i}{\sigma^2_0}\exp(-\frac{w^2_i}{2\sigma^2_0})-\frac{Bw_i}{\sigma^2}\exp(-\frac{w^2_i}{2\sigma^2})]^2}{A\exp(-\frac{w^2_i}{2\sigma^2_0})+B\exp(-\frac{w^2_i}{2\sigma^2})}dw_i
\end{equation}
where the above integral can be decomposed to three integrals
which are
$D_1=\int_{-\infty}^{+\infty}\frac{\frac{-A^2w^2_i}{\sigma^4_0}\exp(-\frac{w^2_i}{2\sigma^2_0})}{A\exp(-\frac{w^2_i}{2\sigma^2_0})+B\exp(-\frac{w^2_i}{2\sigma^2})}dw_i$,
$D_2=\int_{-\infty}^{+\infty}\frac{\frac{-ABw^2_i}{\sigma^2_0\sigma^2}\exp(-\frac{w^2_i}{2\sigma^2_0}-\frac{w^2_i}{2\sigma^2})}{A\exp(-\frac{w^2_i}{2\sigma^2_0})+B\exp(-\frac{w^2_i}{2\sigma^2})}dw_i$
and
$D_3=\int_{-\infty}^{+\infty}\frac{\frac{-B^2w^2_i}{\sigma^4}\exp(-\frac{w^2_i}{2\sigma^2})}{A\exp(-\frac{w^2_i}{2\sigma^2_0})+B\exp(-\frac{w^2_i}{2\sigma^2})}dw_i$.
Since we have a term $w^2_i$ in the numerator of the above
integrals and the Gaussian term with small variance is large near
zero, we can neglect the Gaussian term with small variance (delta
function) in the denominator. So, the integrals $D_1$ and $D_2$
with neglecting this term will be approximately zero. We verify
this approximation in the simulation results by computing these
integrals numerically. Finally, the third integral will be
approximately
$D_3\approx\int_{-\infty}^{+\infty}\frac{\frac{-B^2w^2_i}{\sigma^4}\exp(-\frac{w^2_i}{2\sigma^2})}{B\exp(-\frac{w^2_i}{2\sigma^2})}dw_i$.
Calculating this integral results is $J_{P_{ii}}\approx
D_3\approx\frac{1-p}{\sigma^2}$.


\end{document}